# From Job Postings to Curriculum Decisions:

# Using AI to Generate Workforce Intelligence for MSW Program Planning

Barbara S. Hiltz[1], Bryan G. Victor[2], Brian E. Perron[1]

[1]School of Social Work, University of Michigan

[2]School of Social Work, Wayne State University

## Author Note

Bryan G. Victor: https://orcid.org/0000-0002-2092-912X

Brian E. Perron: https://orcid.org/0009-0008-4865-451X

We have no known conflict of interest to disclose.

Correspondence concerning this article should be addressed to Brian E. Perron, University of Michigan School of Social Work, 1080 S. University Avenue, Ann Arbor, MI 48109.

beperron@umich.edu

## Abstract

Social work programs lack systematic methods to align curricula with employer expectations, typically relying on advisory input and alumni surveys rather than direct analysis of workforce requirements. This paper presents a case study demonstrating how one MSW program used artificial intelligence tools to generate organizational intelligence from job posting data for curriculum planning. Using a locally deployed language model, we classified over 40,000 job postings for MSW relevance and alignment with eight practice specializations, then extracted skills, therapeutic modalities, and technology competencies. Interpersonal Practice dominated the employment landscape, followed by Children, Youth, and Families. Clinical Assessment and Case Management emerged as cross-cutting competencies. Macro-level specializations showed co-occurrence patterns among partially aligned positions that largely disappeared among positions requiring MSW credentials specifically. Trauma-informed care appeared in management and evaluation roles, reflecting its expansion from clinical modality to organizational framework. The methodology demonstrates a transferable approach that other programs can adapt for strategic planning, and the findings illustrate the type of intelligence such analysis can yield. The patterns identified entered faculty deliberation as one input among many, interpreted by stakeholders with contextual knowledge no dataset can fully capture.

## From Job Postings to Curriculum Decisions: Using AI to Generate Workforce Intelligence for MSW Program Planning

Social work programs face a persistent challenge in aligning their curricula with workforce demands in the absence of systematic, current data on employer-specified skills. Programs operating under the Council on Social Work Education's competency-based accreditation model, as articulated in the 2022 Educational Policy and Accreditation Standards (EPAS), have wide discretion in determining how to operationalize the nine core competencies. Yet curriculum decisions typically rely on advisory board input, alumni surveys, and anecdotal faculty and employer feedback rather than direct analysis of employer requirements. The sources currently used by programs are valuable but constitute limited samples and retrospective self-reports rather than current, employer-stated expectations.

This paper presents a case study of an MSW program that used artificial intelligence (AI) tools to curate and analyze large-scale job-posting data to inform curriculum planning. A distinction is necessary at the outset. This project was designed to generate organizational intelligence for curriculum decisions, defined here as the systematic use of timely, institution-specific data to inform internal deliberation and planning. Organizational decision-making operates under different evidentiary standards than scientific knowledge production. The relevant question is not whether findings would replicate across all MSW programs, but whether the evidence provides sufficient confidence to inform curriculum deliberation at this institution. This distinction matters: the methods described below were designed to serve a specific institutional planning process, and the findings are presented as illustrative of the intelligence such an analysis can yield rather than as generalizable claims about the social work labor market. Other programs applying this methodology would generate institution-specific results reflecting their own curricular frameworks, geographic labor markets, and strategic priorities. The patterns identified entered faculty discussion as one input among many, interpreted by stakeholders with contextual knowledge that no dataset can fully capture. The remainder of this paper describes the context motivating this analysis, the methods employed, the patterns observed, and their implications for curriculum planning.

## Context: The Challenge of Workforce Alignment

The scale and diversity of social work employment complicate curriculum-workforce alignment. The 2024 Social Work Workforce Study Series (Association of Social Work Boards [ASWB], 2025b) estimated approximately 463,000 licensed social workers in the United States. Social work employment is projected to grow 6% through 2034, faster than average for all occupations (Bureau of Labor Statistics, 2025). This growth spans mental health services, substance use disorder treatment, aging-related services, child welfare, and numerous other domains - each with potentially distinct skill requirements that programs must anticipate.

The most systematically collected evidence on practice competencies comes from the ASWB's 2024 Analysis of the Practice of Social Work (ASWB, 2025a), based on survey data from over 25,000 licensed and registered social workers. The practice analysis identified common elements of practice and synthesized them into knowledge, skills, and abilities (KSA) frameworks at various licensure levels. The 2024 revision consolidated content into three domains - Values and Ethics, Assessment and Planning, and Intervention and Practice - with substantial overlap across practice levels. Drawing on analysis of practice data, Apgar (2025) found 97% content overlap between clinical and advanced generalist examinations, prompting a call for universal advanced generalist education rather than specialized tracks.

While the KSA framework offers valuable insight into shared professional practice, it has limitations as a basis for curriculum planning. The practice analysis captures what practitioners report doing across their careers rather than the criteria employers specify when hiring for particular roles. Aggregated self-report data obscure variation in skill expectations across settings, positions, and career stages - particularly at the workforce entry stage. Moreover, the analysis does not capture how employers signal required competencies through job descriptions and hiring requirements.

**Job Postings as a Source of Organizational Intelligence**

Job postings represent a largely untapped data source for understanding employer expectations. These documents capture requirements at the time of hire and constitute explicit statements of the skills, credentials, and competencies that organizations believe are necessary for successful performance. Recent advances in natural language processing have made large-scale analysis of such unstructured text feasible, enabling the extraction of granular skill information from thousands of postings.

A growing body of work demonstrates the value of job posting analysis for organizational intelligence. Rahhal et al. (2024) surveyed data science techniques for job market analysis, documenting methods including named entity recognition, skill taxonomy mapping, and machine learning classification. Within health professions education, Krasna (2024) analyzed 70,343 job postings for Master of Public Health graduates, contrasting employer-desired skills with accreditation competencies. The study found strong overall alignment but noted that some competencies appeared less frequently as explicit hiring criteria - highlighting potential differences between foundational competencies employers assume and those they explicitly signal.

To our knowledge, no comparable analysis exists for social work. This gap motivated the present project. We aimed to develop and demonstrate a methodology for extracting workforce intelligence from job postings that could inform curriculum decisions at the institution-level and, potentially, serve as a model for other programs.

### Guiding Questions

Three questions guided this analysis:

1. What is the distribution of workforce demand across the institution's MSW practice specializations, and how do positions cluster across multiple specializations?
2. What specific clinical, technical, and software skills do employers require for positions seeking to hire MSW-prepared professionals?
3. What evidence-based therapeutic modalities appear most frequently in employer requirements MSW-relevant positions?

These questions were formulated to address a specific curriculum initiative at an MSW program that organizes its MSW program around eight practice specializations. The analysis maps workforce demand onto this institutional framework, providing intelligence directly applicable to program decisions. Programs with different specialization structures would adapt the methodology accordingly.

## Methods

## Data Collection

Job postings were collected from three major employment platforms (Indeed, LinkedIn, and Glassdoor) using Octoparse (Octopus Data Inc., 2016), a no-code web scraping tool. A comprehensive search strategy employed 79 distinct search terms spanning clinical practice (e.g., "licensed clinical social worker," "therapist," "crisis counselor"), child welfare (e.g., "child protective services," "foster care case manager"), program evaluation (e.g., "program evaluator," "research coordinator"), and administrative roles (e.g., "nonprofit manager," "program director"). The complete set of search terms is included in Appendix A. Data extraction was performed throughout December 2025, yielding 41,584 unique job postings across all 50 U.S. states and territories with no geographic filtering applied. The dataset represents a national cross-sectional snapshot of positions available during the collection period rather than a longitudinal sample of hiring patterns.

The data collection and processing approach extends methods demonstrated in prior social work research using language models to classify and extract information from unstructured text (Perron et al., 2025). Like other systematic analyses of job postings, this project addressed common challenges, including data quality assurance, entity normalization (standardizing job titles, locations, and skill terms), and deduplication across multiple sources (Tzimas et al., 2024).

## Classification Procedures

Classification and extraction tasks were performed using a 20-billion-parameter language model, GPT-OSS:20B (OpenAI, 2025), deployed on local computing hardware via the llama.cpp inference engine (Gerganov, 2023). The classification task requires assigning unstructured natural-language text - full job descriptions typically spanning 300 to 500 words - to nuanced, institution-specific categories. Traditional supervised machine learning approaches (e.g., support vector machines, random forests, naive Bayes classifiers) would require large manually labeled training datasets for each of the eight specializations, feature engineering to convert free-text descriptions into numerical representations, and separate model training and validation for each classification task. No labeled training corpus existed for this institution-specific classification scheme, and creating one would itself require the expert coding effort the LLM-based approach was designed to supplement.

LLMs perform zero-shot classification through structured prompts, eliminating these prerequisites while leveraging the model’s pre-trained understanding of natural language. The task parameters - unstructured text, institution-specific categories, no existing labeled data - align with LLM strengths. This model was additionally selected for its computational efficiency and reduced environmental impact relative to large commercial models such as ChatGPT, Claude, and Gemini. Recent research demonstrates that smaller language models achieve comparable performance to larger models on classification tasks while consuming substantially fewer computational resources (Qi et al., 2026).

Each job posting was classified along two dimensions: its relevance to MSW graduates and its alignment with specific MSW practice specializations. Skills and competencies mentioned in each posting were then extracted and categorized. Classification prompts, pipeline documentation, data, and the complete set of search terms are available from the corresponding author upon request. Full job posting data cannot be shared publicly due to the terms of service of the source platforms, which restrict the redistribution of scraped content.

***MSW Relevance Screening.*** Each posting was evaluated for alignment with MSW education and practice and assigned to one of three categories:

- Strong alignment: The position explicitly requires or strongly prefers an MSW degree or MSW-level licensure (LCSW, LMSW, etc.), or core responsibilities directly correspond to social work practice domains, including clinical services, case management, discharge planning, community practice, social service administration, policy advocacy, or clinical supervision.
- Partial alignment: The position involves human services work where MSW training is applicable but not central. This includes roles where social work is one of several accepted credentials, where responsibilities overlap with but are not primarily social work functions, or where the setting employs social workers but the specific role is adjacent (e.g., care coordination, victim advocacy, health education).
- No alignment: The position requires credentials from distinct professions without a social work degree, involves minimal human services content, or operates in domains outside the social work scope.

***Specialization Classification.*** Positions meeting the relevance threshold (strong or partial alignment) were independently evaluated against the eight MSW practice specializations that constitute the program's published curriculum framework: Interpersonal Practice, Children/Youth/Families, Management/Leadership, Older Adults, Program Evaluation/Applied Research, Community Change, Policy/Political Social

Work, and Global Social Work. These are the program's existing specializations, as defined in its academic catalog - not an analytical taxonomy created for this study. "Interpersonal Practice," for instance, is this program's name for its clinical specilaization encompassing mental health practice; "Management and Leadership" is a distinct specialization within the curriculum. The study asks: given our program's existing structure, what does the job market look like? Programs with different specialization structures would map their own curricular frameworks using the same methodology, substituting their own definitions for the classification prompts. Detailed definitions for each specialization appear in Appendix B.

Classification prompts were structured around three components: core indicators characteristic of each specialization, typical employment settings, and explicit decision rules for classification boundaries. For example, the Interpersonal Practice prompt specified clinical social work, psychotherapy, evidence-based interventions (cognitive behavioral therapy, dialectical behavioral therapy, motivational interviewing), mental health assessment, and clinical supervision as core indicators, with community mental health centers, hospitals, and outpatient clinics as typical settings.

Each position received a classification for all eight specializations, enabling simultaneous identification of positions aligned with multiple specializations. This multi-label approach reflects the reality that many social work positions require competencies spanning traditional specialization boundaries.

***Skills Extraction.*** Four categories of skills were extracted from each posting:

1. Therapeutic modalities: Named therapeutic approaches such as CBT, DBT, Motivational Interviewing, or Trauma-Informed Care

2. Technical skills: Professional competencies requiring specialized training, such as clinical assessment, discharge planning, grant writing, or program evaluation
3. Soft skills: Interpersonal and transferable competencies such as communication, teamwork, or cultural competence
4. Technology skills: Software platforms and digital competencies including Electronic Health Record systems, statistical software, and office applications

Each extracted skill was tagged with a requirement level (required versus preferred) based on the language in the job posting. Following extraction, extensive normalization was applied to standardize terminology. Synonymous terms were mapped to canonical forms (e.g., "CBT," "cognitive behavioral," and "cognitive-behavioral therapy" were consolidated as "Cognitive Behavioral Therapy").

**Quality Assurance**

The relevant quality question for this project was whether the classification and extraction processes produced patterns reliable enough to inform curriculum deliberation. We approached this as an organizational quality assurance task rather than a scientific validation exercise.

***Classification Review.*** We developed classification prompts iteratively, generating condensed summaries of job postings that retained substantive content (responsibilities, credentials, practice setting) while omitting extraneous material (benefits, salary, application procedures). These summaries enabled rapid review during prompt refinement.

To confirm that the final classification process aligned with professional judgment, a subject matter expert reviewed a sample of postings alongside the model-generated rationales. The expert's classifications corresponded with model output in the substantial majority of cases for both MSW relevance screening and specialization assignment. Disagreements occurred in predictable edge cases - positions with ambiguous credential requirements or hybrid responsibilities spanning specialization boundaries - rather than systematic errors that would distort aggregate patterns.

Approximately 7.7% of MSW-relevant postings were not assigned to any specialization. Review of a sample suggested the model was conservative; most could reasonably have been assigned to one or more specializations. This pattern results in modest undercounting within specializations rather than misclassification, which we judged acceptable for the purposes of identifying workforce patterns.

***Extraction Review.*** Skills extraction was reviewed through both manual inspection and automated evaluation using a separate language model to compare extracted skills against original job descriptions. The extraction process reliably captured skills mentioned in postings, with strongest performance for therapeutic modalities and named software systems that appear as distinct terms. Performance was weaker for soft skills and generic competencies embedded in prose rather than listed explicitly. For aggregate analysis of skill frequency across specializations, this performance was sufficient. The goal was to surface prevalent patterns, not to achieve the posting-level precision that applications such as automated job matching would require.

***Summary.*** The quality assurance procedures confirmed that classification and extraction produce results consistent with professional judgment and reliable enough to support pattern interpretation. Any large-scale automated analysis will contain errors at the individual posting level; the relevant standard is whether aggregate patterns reflect workforce conditions with sufficient fidelity to warrant faculty deliberation. We concluded that they do.

**Analysis**

Descriptive statistics, including counts, proportions, and market share percentages, were calculated for specialization demand distribution. Market share was calculated as the number of positions aligned with each specialization divided by total MSW-aligned positions. Because positions could align with multiple specializations, market shares sum to greater than 100%.

To examine patterns of specialization co-occurrence - the extent to which job postings aligned with multiple MSW specializations simultaneously - we constructed a binary alignment matrix, coding each position as aligned or not aligned for each specialization. The matrix was constructed by the first and third authors, who assessed each specialization definition against the classification prompts used by the language model. Because the matrix maps the program's own published curriculum descriptions to LLM classification prompts - a task primarily requiring accurate translation of institutional definitions rather than subjective interpretation - disagreements were infrequent and were resolved through discussion and reference to the published specialization descriptions. We then computed pairwise phi coefficients ($\varphi$) to quantify the association between specialization alignments across all positions. The phi coefficient is a measure of association for dichotomous variables, ranging from −1 (perfect negative association) to +1 (perfect positive association), with values near zero indicating no systematic relationship. Positive coefficients indicate specializations that frequently appear together in job requirements; negative coefficients indicate specializations that rarely co-occur within the same position.

This exploratory analysis identified specialization clusters that may reflect employer expectations for integrated competencies spanning traditional boundaries - patterns relevant to curriculum design decisions about specialization structure. For skills analysis, we calculated frequency distributions and requirement levels (required versus preferred) for technical skills, therapeutic modalities, and technology skills within each specialization. The top five skills for each specialization were identified by computing the proportion of positions in that specialization that mentioned each skill.

## Results

### Specialization Demand Distribution

Of the 41,584 deduplicated job postings, 23,732 (57.1%) were retained for analysis based on MSW relevance screening. This included 7,791 positions classified as strongly aligned with MSW education and practice and 15,941 classified as partially aligned. Figure 1 presents the distribution of retained positions across the eight MSW practice specializations. Because positions are frequently aligned with multiple specializations, market share percentages sum to greater than 100%. The findings presented below represent patterns observed in this dataset during the collection period. For organizational intelligence purposes, the value lies in the relative distribution across specializations and the skills that emerge as prevalent, not in the precise percentages, which would vary with different collection periods, platforms, or search strategies. Readers should interpret specific figures as indicative of patterns rather than as precise population estimates.

[INSERT FIGURE 1 ABOUT HERE]

Interpersonal Practice dominated the employment landscape, with 16,597 positions (69.9% market share). The strong alignment classification identified 4,811 Interpersonal Practice positions explicitly requiring clinical credentials or involving clinical practice responsibilities. Children, Youth, and Families represented the second-largest demand area, comprising over 40% of aligned positions, with positions including child welfare case managers, foster care coordinators, family preservation specialists, and youth program directors. The high rate for strong alignment (4,263 positions) indicates many employers specifically seek MSW credentials for child welfare roles.

Management and Leadership (5,363 positions; 22.6%) demonstrated that supervisory and administrative roles constitute a substantial segment of the MSW job market, typically requiring post-MSW experience. Older adult positions (5,314; 22.4%) represented a substantial portion of the market, including roles in skilled nursing facilities, home health care, geriatric care management, hospice services, and senior centers. The macro-level specializations - Community Change (11.3%), Policy/Political (3.6%), and Global Social Work (3.3%) - represented smaller but significant segments. While numerically fewer in number, these positions often involve high-impact work in advocacy, program development, and systems change.

**Specialization Co-occurrence Patterns**

Job postings frequently aligned with multiple specializations. Figure 2 displays the correlation matrix of phi coefficients for specialization co-occurrence. The lower triangle presents coefficients for all aligned positions (n = 23,732), combining strongly and partially aligned classifications; the upper triangle presents coefficients for strongly aligned positions only (n = 7,791).

[INSERT FIGURE 2 ABOUT HERE]

Two distinct patterns emerged depending on the alignment level. Among all aligned positions, macro-level specializations formed a clear cluster: Management and Leadership, Program Evaluation and Research, Community Change, and Policy and Political showed moderate-to-strong positive correlations among themselves. The strongest associations appeared between Management and Leadership–Program Evaluation and Research and Community Change–Policy and Political. This clustering largely disappeared among strongly aligned positions, where correlations among macro-level specializations fell to near zero.

The negative correlation between Children, Youth, and Families and Older Adults persisted across both alignment levels, reflecting population-specific hiring. The modest positive correlation between Interpersonal Practice and Older Adults among all aligned positions did not persist among strongly aligned positions.

**Skills Analysis**

The skills analysis below focuses on strongly aligned positions - those explicitly requiring or strongly preferring MSW credentials. This restriction produces a clearer signal for curriculum planning: skill requirements in these postings reflect what employers expect specifically from MSW graduates. Including partially aligned positions, which accept an MSW among several credentials, would dilute the signal with requirements from adjacent fields such as counseling, public health, or public administration, where competency expectations may differ from those directed at MSW-prepared professionals.

***Technical and Clinical Skills.*** Technical skills represent core competencies employers specify for MSW positions. Table 1 presents the top five technical skills for each MSW specialization among strongly aligned positions, revealing both cross-cutting competencies and specialization-specific skill demands.

[INSERT TABLE 1 ABOUT HERE]

Clinical Assessment and Case Management emerged as the most prevalent competencies across specializations. Clinical Assessment ranked as the top-ranked skill for Interpersonal Practice (30% of positions) and Older Adults (28%), and among the top three for Children, Youth, and Families (8%) and Global Social Work (13%). Case Management showed similar breadth, ranking first for Children, Youth, and Families (31%) and Global Social Work (37%), second for Interpersonal Practice (16%), and third for Older Adults (18%). This cross-specialization prevalence validates the centrality of assessment and coordination competencies in MSW foundation curricula.

Specialization-specific skill profiles emerged clearly in the data. Clinical specializations emphasized direct practice competencies: Discharge Planning appeared prominently in both Interpersonal Practice (11%) and Older Adults (18%), while Crisis Intervention ranked among the top five skills for Interpersonal Practice (8%), Children, Youth, and Families (7%), and Global Social Work (13%). Macro-level specializations showed distinct competency demands: Program Evaluation ranked first for both Management and Leadership (17%) and Program Evaluation and Research (28%). In comparison, Project Management appeared as the top skill for both Community Change (9%) and Policy and Political (15%). Data Analysis concentrated in research-oriented positions, appearing in Management and Leadership (11%) and Program Evaluation and Research (26%).

***Evidence-Based Therapeutic Modalities.*** Explicit mention of evidence-based therapeutic modalities occurred infrequently across job postings. As summarized in Table 2, Group therapy (28.4%) and cognitive behavioral therapy (5.1%) cluster within Interpersonal Practice positions, while crisis intervention (24.2%) and trauma-informed care (22.6%) appear most frequently in Global Social Work postings - consistent with the prevalence of refugee services and displacement-related trauma work in this specialization. Wraparound services and multisystemic therapy concentrate in Children, Youth, and Families positions, reflecting these modalities' origins in child welfare and juvenile justice systems.

Notably, trauma-informed care appeared among the top modalities in Management and Leadership (11.1%), Program Evaluation and Research (5.7%), and Policy and Political (1.4%) specializations. These positions are not traditionally associated with direct clinical intervention. A manual review of the underlying job descriptions confirmed these were not classification errors but rather reflected the expansion of trauma-informed care from a clinical modality to an organizational and systems-level framework. Management positions referenced trauma-informed care in the context of building organizational cultures, supervising clinical staff, and ensuring program fidelity. Program evaluation roles described assessing trauma-informed service delivery and monitoring the quality of implementation. Policy positions referenced the development of trauma-informed policy frameworks and systems-level interventions.

[INSERT TABLE 2 ABOUT HERE]

***Technology Skills.*** Technology skill requirements varied across MSW specializations (see Table 3). Microsoft Office appeared in the top five skills for seven specializations, with mention rates ranging from 7% in Interpersonal Practice to 57% in Global Social Work. Microsoft Excel appeared separately in all eight specializations, with the highest rates in Global Social Work (37%), Program Evaluation and Research (27%), and Policy and Political (23%). Clinical specializations showed distinct technology profiles. Interpersonal Practice positions most frequently mentioned HIPAA Compliance (9%), Electronic Health Records (6%), and Clinical Documentation (6%). Electronic Health Records also appeared in Older Adults (5%), Children, Youth, and Families (3%), and Management and Leadership (3%). Database Management appeared in six specializations, with rates ranging from 7% in Policy and Political to 31% in Global Social Work. Google Workspace appeared in Community Change (21%), Policy and Political (7%), and Global Social Work (5%).

The Program Evaluation and Research specialization showed the highest concentration of data-related technology requirements. Statistical software appeared in 32% of positions, with specific packages including SPSS (13.7%), SAS (12.3%), Stata (9.0%), and R/RStudio (2.4%). Data visualization tools were mentioned in 20% of positions, with Power BI (9.9%) and Tableau (9.4%) most frequently. Quantitative data platforms included SQL (4.7%) and Qualtrics (2.8%). Qualitative analysis software appeared less regularly, with NVivo (2.8%) and ATLAS.ti (0.9%).

[INSERT TABLE 3 ABOUT HERE]

## Discussion

The distribution of positions across specializations provides a snapshot of employer demand during the collection period. Interpersonal Practice's dominant market share, representing roughly 70% of aligned positions, likely reflects the ongoing mental health workforce shortage and the expansion of insurance coverage for behavioral health services following the Affordable Care Act's mental health parity provisions (Council of State Governments, 2024). The strong alignment rate for Interpersonal Practice (4,811 positions explicitly requiring clinical credentials) suggests that many employers distinguish between MSW-prepared clinicians and those with related qualifications - a pattern with implications for how programs position clinical training within their curricula. Children, Youth, and Families represented the second-largest demand area (42.8%), with the high strong-alignment rate (4,263 positions) indicating that child welfare employers frequently seek MSW credentials specifically. This pattern is consistent with Title IV-E funding requirements and state licensing standards that privilege social work education for child welfare practice.

The macro-level specializations - Community Change (11.3%), Policy and Political (3.6%), and Global Social Work (3.3%) - were numerically smaller. These differences should not be interpreted as indicating lesser importance; macro-level positions often involve high-impact work in advocacy, systems change, and program development. The smaller numbers partly reflect organizational structure:

administrative, supervisory, and leadership tiers are inherently narrower than frontline service delivery, so fewer such positions exist regardless of demand for macro competencies. The smaller numbers may also reflect that macro positions are more likely to be posted through professional networks, organizational websites, or specialized job boards not captured in this analysis. For curriculum planning purposes, the question is not simply how many positions exist but what competencies those positions require and how programs can prepare students for them.

### Specialization Clustering and Curriculum Structure

The correlation analysis revealed patterns that may inform decisions about the structure of specialization. Among all aligned positions, macro-level specializations formed a cluster: Management and Leadership, Program Evaluation and Research, Community Change, and Policy and Political showed moderate-to-strong positive correlations ($\varphi$ = 0.25 to 0.44). This clustering largely disappeared among strongly aligned positions, where correlations fell to near zero ($\varphi$ = −0.01 to 0.12).

This divergence carries interpretive implications worth considering. Positions that partially align with MSW training - those accepting related credentials alongside social work - tend to bundle macro competencies together, seeking practitioners who can manage programs while evaluating them, or organize communities while advocating for policy change. Positions strongly aligned with MSW training tend toward greater specialization. One interpretation is that the MSW credential may serve as a marker of specialized depth in the macro domain: employers willing to accept related degrees seek generalist macro competence, while employers requiring MSW credentials seek specialization-specific expertise.

For curriculum planning, this pattern might support both integrated macro training (preparing students for the broader market of partially aligned positions) and opportunities for specialized depth (for students targeting positions requiring social work credentials specifically). Whether this interpretation holds and how it should inform curriculum design are questions for faculty deliberation informed by, but not determined by, these data.

The persistent negative correlation between Children, Youth, and Families and Older Adults ($\varphi = -0.27$ among all aligned; $\varphi = -0.10$ among strongly aligned) reflects population-specific hiring patterns. Skills in child development, family systems assessment, and permanency planning do not readily transfer to geriatric care management, dementia assessment, or end-of-life counseling. This pattern suggests value in maintaining distinct population-focused training rather than assuming clinical skills transfer automatically across the lifespan - though again, this is an inference from workforce data that warrants consideration alongside other factors in curriculum design.

**Clinical Assessment as a Cross-Cutting Competency**

Clinical Assessment emerged as the most frequently mentioned technical competency across specializations, ranking among the top three skills for Children, Youth, and Families (8%) and Global Social Work (13%), and as the top-ranked skill for Interpersonal Practice (30%) and Older Adults (28%). This cross-specialization prevalence suggests that assessment skills may warrant treatment as foundational competencies applicable to all students rather than as clinical specialization content alone.

Case Management showed similar breadth, ranking first for Children, Youth, and Families (31%) and Global Social Work (37%), second for Interpersonal Practice (16%), and third for Older Adults (18%). The coordination function inherent to case management - navigating complex service systems, linking clients to resources, monitoring progress across providers - appears central to social work practice across populations and settings. Contemporary case management requires navigating healthcare, housing, benefits, legal, and immigration systems, suggesting that case management curriculum might extend beyond generic coordination skills to include system-specific knowledge.

**The Paradox of Therapeutic Modality Specification**

The findings regarding evidence-based therapeutic modalities present an apparent paradox worth noting. Even the most commonly cited approaches (i.e., crisis intervention and group therapy) appeared in fewer than one in four strongly aligned positions. The specific frequencies matter less than the overall pattern: explicit modality specification was uncommon, but when employers did specify modalities, they treated them as requirements rather than preferences. Cognitive Behavioral Therapy, despite its prominence in clinical training curricula, appeared in only 289 postings (2.6% of the sample). This low frequency could suggest that therapeutic modalities are less important to employers than commonly assumed. Alternatively, employers may assume therapeutic competence as a baseline expectation for clinical positions, explicitly specifying modalities only when particular approaches are essential. This finding illustrates precisely why AI-generated workforce data should supplement, not replace, other evidence sources. If a program relied solely on job posting analysis, it might erroneously conclude that CBT training is less important than its actual centrality to clinical practice warrants--an inference contradicted by clinical training standards, licensing examination content, and practice wisdom. The low frequency of CBT in postings likely reflects employer assumptions about baseline competence rather than indifference to the modality. This concrete example underscores the paper's broader argument: workforce intelligence functions best as one input among several in curriculum deliberation, and patterns that seem surprising in isolation often resolve when triangulated with other sources of evidence.

**Trauma-Informed Care Across Practice Levels**

A notable observation concerns the appearance of Trauma-Informed Care in positions not traditionally associated with direct clinical intervention. Trauma-Informed Care appeared among the top modalities in Management and Leadership (11.1%), Program Evaluation and Research (5.7%), and Policy and Political (1.4%) specializations. Manual review of the underlying job descriptions confirmed these were not classification errors but reflected the expansion of Trauma-Informed Care from a clinical modality to an organizational and systems-level framework.

Management positions referenced Trauma-Informed Care in the context of building organizational cultures, supervising clinical staff, and ensuring program fidelity. Program evaluation roles described assessing trauma-informed service delivery and monitoring implementation quality. Policy positions referenced trauma-informed policy frameworks and systems-level interventions. This pattern suggests that contemporary employers may expect macro-level practitioners to understand trauma-informed principles - not to deliver individual therapy, but to create organizational environments, evaluation protocols, and policies that support trauma-informed service delivery. The finding supports considering whether trauma-informed content should permeate the curriculum rather than be confined to clinical courses.

**Technology Competencies Across Specializations**

Technology skill requirements varied meaningfully across specializations. Clinically-oriented specializations showed distinct technology profiles centered on HIPAA Compliance, Electronic Health Records (EHR), and Clinical Documentation. The concentration of EHR requirements in clinical specializations carries implications for field preparation: healthcare settings increasingly require EHR proficiency. While specific systems vary across sites, foundational training in electronic documentation principles and health information privacy would prepare students for employment where such competence is expected.

The Program Evaluation and Research specialization showed the highest concentration of data-related technology requirements, with statistical software appearing in roughly one-third of positions - substantially higher than in any other specialization. This concentration suggests that students entering evaluation-focused careers need substantive training in quantitative tools. This is a finding that aligns with longstanding discussions about research methods preparation in MSW education.

**Considerations for Curriculum Planning**

The patterns identified in this analysis suggest several directions worth considering in curriculum deliberation. These recommendations emerge from observed workforce patterns. The goal is to surface systematic signals that might otherwise remain invisible, not to prescribe specific curricular allocations based on exact market share percentages. These are strategic options informed by systematic examination of employer requirements. Faculty judgment, institutional context, and factors beyond workforce data must shape actual curriculum decisions.

First, the cross-specialization prevalence of Clinical Assessment suggests that strengthening assessment training across all specializations is preferable to confining it to clinical tracks. Foundation courses might provide assessment competencies applicable to all students, with advanced courses deepening specialization-specific applications.

Second, while the frequency of specific therapeutic interventions in job requirements was low, when mentioned, CBT and Motivational Interviewing represent high-value training investments. Trauma-Informed Care was more frequent across specializations. Programs with limited clinical elective capacity might prioritize these approaches, while programs with broader offerings might ensure depth in all three. The expansion of Trauma-Informed Care into macro specializations suggests this content might benefit students across the curriculum.

Third, the technology requirements observed across specializations - EHR systems for clinical practice, statistical software for evaluation, productivity tools throughout - suggest value in integrating technology training into field preparation and coursework.

Fourth, the strong correlations among macro-level specializations raise questions about whether maintaining separate tracks for Management and Leadership, Program Evaluation, Community Change, and Policy and Political reflects workforce expectations or creates artificial divisions. Programs might consider tracks combining related competencies - for example, management with evaluation, community organizing with policy analysis - while recognizing that strongly aligned positions often seek specialized expertise.

Our findings engage with recent calls for restructuring MSW education. Apgar (2025), drawing on the ASWB practice analysis, argued that substantial overlap in knowledge and skills across practice settings supports eliminating clinical and nonclinical specializations in favor of advanced generalist graduate education. The workforce data provide partial support for this position: moderate-to-strong correlations across macro specializations indicate that employers often expect integrated competencies that span traditional boundaries.

However, the clinical domain presents more complex patterns. The strong alignment rate for Interpersonal Practice - 4,811 positions with 60.3% mentioning licensure requirements - indicates that employers distinguish between MSW-prepared clinicians and practitioners with related qualifications. The persistent negative correlation between Children, Youth, and Families and Older Adults reflects population-specific competencies that do not readily transfer across populations. Rather than adopting either rigid specialization or complete generalist training, the patterns observed here suggest that consolidating some macro-level tracks while maintaining population-specific training may warrant consideration.

Fifth, Management and Leadership positions constitute 22.6% of the market, representing a substantial segment typically requiring post-MSW experience. Even students focused on clinical careers may benefit from exposure to supervision, program management, and administrative competencies that support advancement into leadership roles.

**Methodological Contribution**

Beyond the substantive patterns, this project demonstrates a methodology for workforce analysis that other programs can adapt. The three-stage pipeline - relevance screening, specialization classification, and skills extraction - transforms unstructured job posting text into analyzable data using locally deployed language models that preserve data privacy and minimize processing costs.

This approach offers advantages over traditional methods for gathering workforce intelligence. The scale of 23,732 analyzed positions far exceeds typical survey samples. Direct extraction from employer-written text captures requirements at the moment of hiring rather than retrospective self-reports. Individual skill extraction enables specific rather than generic findings. Local deployment means any program can replicate the methodology without commercial analytics contracts or specialized data science expertise.

The use of smaller, efficient language models makes this approach accessible. The 20-billion-parameter model achieved classification accuracy sufficient for organizational intelligence purposes while consuming substantially fewer computational resources than large commercial models. This efficiency democratizes workforce analytics: programs can conduct ongoing monitoring of local labor markets without large technology budgets. If multiple programs conducted parallel analyses using their own specialization structures, the resulting cross-institutional comparisons could contribute to a more systematic understanding of how social work education maps onto workforce expectations - and might ultimately inform the kind of standardized competency definitions that would benefit the profession as a whole.

**Scope and Limitations**

Several characteristics define the scope of this analysis. These reflect deliberate choices appropriate for organizational intelligence rather than limitations requiring mitigation.

The dataset represents a cross-sectional snapshot from a single month. For scientific research purposes, this would constrain conclusions about trends or causal relationships. For curriculum planning, a snapshot of current employer requirements provides actionable intelligence; curriculum decisions operate on multi-year cycles and benefit from systematic examination of present conditions rather than longitudinal precision.

Job postings capture employer aspirations at the hiring stage rather than actual hiring decisions or job performance requirements. What employers say they want may differ from whom they ultimately hire or which skills prove most valuable in practice. This gap between stated requirements and realized practice is inherent in job posting data. For curriculum planning, employer-stated requirements constitute one meaningful signal about workforce expectations, useful alongside other sources of information.

Geographic representation reflects platform posting patterns. Positions posted only on agency websites or through professional networks were not captured. The three platforms analyzed likely overrepresent larger employers and geographic areas with active online job markets. Programs serving particular geographic markets or student populations might conduct targeted analyses of their specific labor markets rather than relying on national patterns.

The specialization framework derives from this program's curriculum and may not align with structures at other institutions. Programs adapting this methodology would map findings to their own curricular frameworks. The prompts, classification criteria, and analytical approach are transferable; the specific specialization definitions require local adaptation. Any large-scale automated analysis will contain classification errors at the individual posting level. The quality assurance procedures established that aggregate patterns are reliable enough to support interpretation and deliberation, which is the relevant standard for organizational intelligence.

Several additional characteristics of the data warrant acknowledgment. The three platforms analyzed - Indeed, LinkedIn, and Glassdoor - likely underrepresent smaller community-based organizations that recruit through professional networks, organizational websites, or discipline-specific boards such as the NASW Career Center. Larger employers with dedicated human resources departments are probably overrepresented. Programs conducting similar analyses might supplement major platform data with postings from field-specific sources relevant to their student populations.

No search term set is exhaustive, and additional terms would likely yield additional relevant postings. However, search terms determine which postings enter the dataset; the language model then classifies based on the full content of each posting, not on the retrieval term. Our 79 search terms include broad terms (e.g., “social work,” “social services,” “human services,” “nonprofit manager”) designed to cast a wide net. A posting for a position such as “development director” that mentions social services in its description would be captured by these terms and classified accordingly. The distinction between search term coverage and content-based classification mitigates, though does not eliminate, concerns about term completeness.

Finally, job postings may understate latent demand for competencies that employers believe are unavailable in the applicant pool--a form of filtered demand. If employers have learned that certain skills are rarely found among applicants, they may omit those requirements from postings, understating actual organizational needs. This filtering mechanism means job posting analysis captures expressed demand rather than total demand, reinforcing the argument that workforce intelligence from postings should be interpreted alongside other evidence sources that may capture unmet or aspirational competency needs.

## Conclusion

This project demonstrates one approach to generating workforce intelligence for curriculum planning using AI tools to analyze large-scale job posting data. The methodology - data collection, classification, extraction, and verification - is transferable to other programs seeking systematic insight into their labor markets.

The substantive findings suggest patterns worth considering in curriculum deliberation: the continued prominence of clinical practice (Interpersonal Practice at 69.9% market share with 4,811 strongly aligned positions), the prevalence of positions requiring competencies across traditional specialization boundaries (moderate-to-strong phi coefficients among macro specializations that largely disappeared among strongly aligned positions), cross-cutting demand for assessment and case management skills (Clinical Assessment appearing in the top three skills for four specializations; Case Management ranking first in two), specific therapeutic modalities that employers treat as essential when mentioned (CBT in 289 postings, with over 90% as requirements rather than preferences), and the expansion of trauma-informed frameworks into macro-level practice (appearing in 11.1% of Management and Leadership and 5.7% of Program Evaluation positions). These patterns do not prescribe curriculum decisions but inform the deliberation that faculty, administrators, and stakeholders must undertake.

The distinction between organizational intelligence and scientific research framed this project from the outset. We did not set out to produce generalizable knowledge claims about social work workforce requirements; rather, we sought to inform curriculum planning at one institution through systematic analysis of employer-stated requirements. The methodology and findings may interest other programs facing similar questions, but the appropriate use of this work is as one input to institutional decision-making rather than as scientific evidence demanding particular conclusions.

As social work education continues to adapt to the 2022 EPAS framework, programs need actionable intelligence about workforce expectations. Job posting analysis offers a feasible, scalable approach to gathering such intelligence - one that complements rather than replaces advisory input, alumni feedback, and faculty expertise. The patterns identified here have already entered curriculum deliberation at our institution; we offer the methodology and findings in the hope that other programs may find them useful for their own strategic planning.

**Table 1.**

*Top Five Technical Skills by MSW Specialization (Strongly Aligned Positions)*

| Specialization | n | #1 Skill | #2 Skill | #3 Skill | #4 Skill | #5 Skill |
|---|---|---|---|---|---|---|
| Interpersonal Practice | 4,811 | Clinical Assessment (30%) | Case Management (16%) | Discharge Planning (11%) | Group Therapy (10%) | Crisis Intervention (8%) |
| Children, Youth, and Families | 4,263 | Case Management (31%) | Clinical Assessment (8%) | Crisis Intervention (7%) | Treatment Planning (6%) | Clinical Documentation (3%) |
| Older Adults | 1,066 | Clinical Assessment (28%) | Discharge Planning (18%) | Case Management (18%) | Care Planning (5%) | Hospice Care (4%) |
| Management and Leadership | 637 | Program Evaluation (17%) | Budget Management (13%) | Data Analysis (11%) | Case Management (9%) | Project Management (8%) |
| Program Evaluation and Research | 212 | Program Evaluation (28%) | Data Analysis (26%) | Project Management (17%) | Data Collection (15%) | Data Visualization (12%) |
| Community Change | 81 | Project Management (9%) | Program Evaluation (6%) | Grant Management (4%) | Community Organizing (4%) | Grant Writing (4%) |

| Specialization | n | #1 Skill | #2 Skill | #3 Skill | #4 Skill | #5 Skill |
|---|---|---|---|---|---|---|
| Policy and Political | 71 | Project Management (15%) | Program Evaluation (13%) | Budget Management (10%) | Policy Analysis (10%) | Strategic Planning (8%) |
| Global Social Work | 62 | Case Management (37%) | Clinical Assessment (13%) | Crisis Intervention (13%) | Database Management (8%) | Community Outreach (6%) |

*Note.* Rows ordered by number of strongly aligned positions. Percentages indicate the proportion of specialization positions specifying the skill.

**Table 2.**

*Evidence-Based Therapeutic Modalities in Strongly Aligned Positions*

| Modality | n | 1st Specialization | 2nd Specialization | 3rd Specialization |
|---|---|---|---|---|
| Crisis Intervention | 1,822 | GSW (24.2%) | IP (19.4%) | CYF (17.3%) |
| Group Therapy | 1,809 | IP (28.4%) | CYF (9.1%) | OA (4.2%) |
| Family Therapy | 1,170 | IP (14.5%) | CYF (10.2%) | GSW (3.2%) |
| Trauma-Informed Care | 710 | GSW (22.6%) | ML (11.1%) | CYF (9.2%) |
| Cognitive Behavioral Therapy | 289 | IP (5.1%) | CYF (0.9%) | OA (0.4%) |
| Motivational Interviewing | 243 | IP (3.3%) | CYF (1.7%) | ML (0.8%) |
| De-escalation Techniques | 160 | GSW (4.8%) | ML (2.5%) | CYF (1.8%) |
| Solution-Focused Brief Therapy | 108 | GSW (1.6%) | PP (1.4%) | IP (1.1%) |
| Wraparound Services | 102 | CYF (1.8%) | GSW (1.6%) | ML (1.4%) |
| Multisystemic Therapy | 70 | PP (1.4%) | CYF (1.0%) | IP (0.5%) |

*Note.* GSW = Global Social Work; IP = Interpersonal Practice; CYF = Children, Youth, and Families; OA = Older Adults; ML = Management and Leadership; PP = Policy and Political Practice. n = total mentions. Percentages represent the proportion of postings within each specialization that mentioned the modality. Crisis Intervention appears in both technical skills and therapeutic modalities analyses; job postings variably frame it as a clinical competency or a named intervention approach. Trauma-Informed Care appeared across all eight specializations; only the top three are displayed here. For the full distribution, including Program Evaluation and Research (5.7%) and Policy and Political (1.4%), see the text discussion under “Evidence-Based Therapeutic Modalities.”

**Table 3.**

*Technology and software skills by specialization.*

| Specialization | n | #1 Skill | #2 Skill | #3 Skill | #4 Skill | #5 Skill |
|---|---|---|---|---|---|---|
| Interpersonal Practice | 4,811 | HIPAA Compliance (9%) | Microsoft Office Suite (7%) | Electronic Health Records (7%) | Documentation Systems (6%) | Microsoft Excel (3%) |
| Children, Youth, and Families | 4,263 | Microsoft Office Suite (15%) | Database Systems (7%) | Microsoft Excel (6%) | Documentation Systems (4%) | Electronic Health Records (3%) |
| Management and Leadership | 637 | Microsoft Office Suite (25%) | Database Systems (13%) | Microsoft Excel (11%) | Case Management Software (4%) | Electronic Health Records (3%) |
| Older Adults | 1,066 | Microsoft Office Suite (9%) | Documentation Systems (6%) | Microsoft Excel (5%) | Electronic Health Records (5%) | HIPAA Compliance (3%) |
| Program Evaluation and Research | 212 | Statistical Software (32%) | Microsoft Excel (27%) | Database Systems (26%) | Microsoft Office Suite (25%) | Telehealth Platforms (5%) |
| Community Change | 81 | Google Workspace (21%) | Database Systems (16%) | Microsoft Office Suite (15%) | Microsoft Excel (10%) | Video Conferencing (4%) |
| Policy and Political | 71 | Microsoft Office Suite (27%) | Microsoft Excel (23%) | Statistical Software (10%) | Database Systems (7%) | Google Workspace (7%) |
| Global Social Work | 62 | Microsoft Office Suite (57%) | Microsoft Excel (37%) | Database Systems (31%) | Documentation Systems (7%) | Telehealth Platforms (5%) |

*Note.* Percentages represent the proportion of strongly aligned postings within each specialization that mentioned the technology skill.

**Figure 1.**

*Distribution of MSW-Aligned Positions Across Practice Specializations*

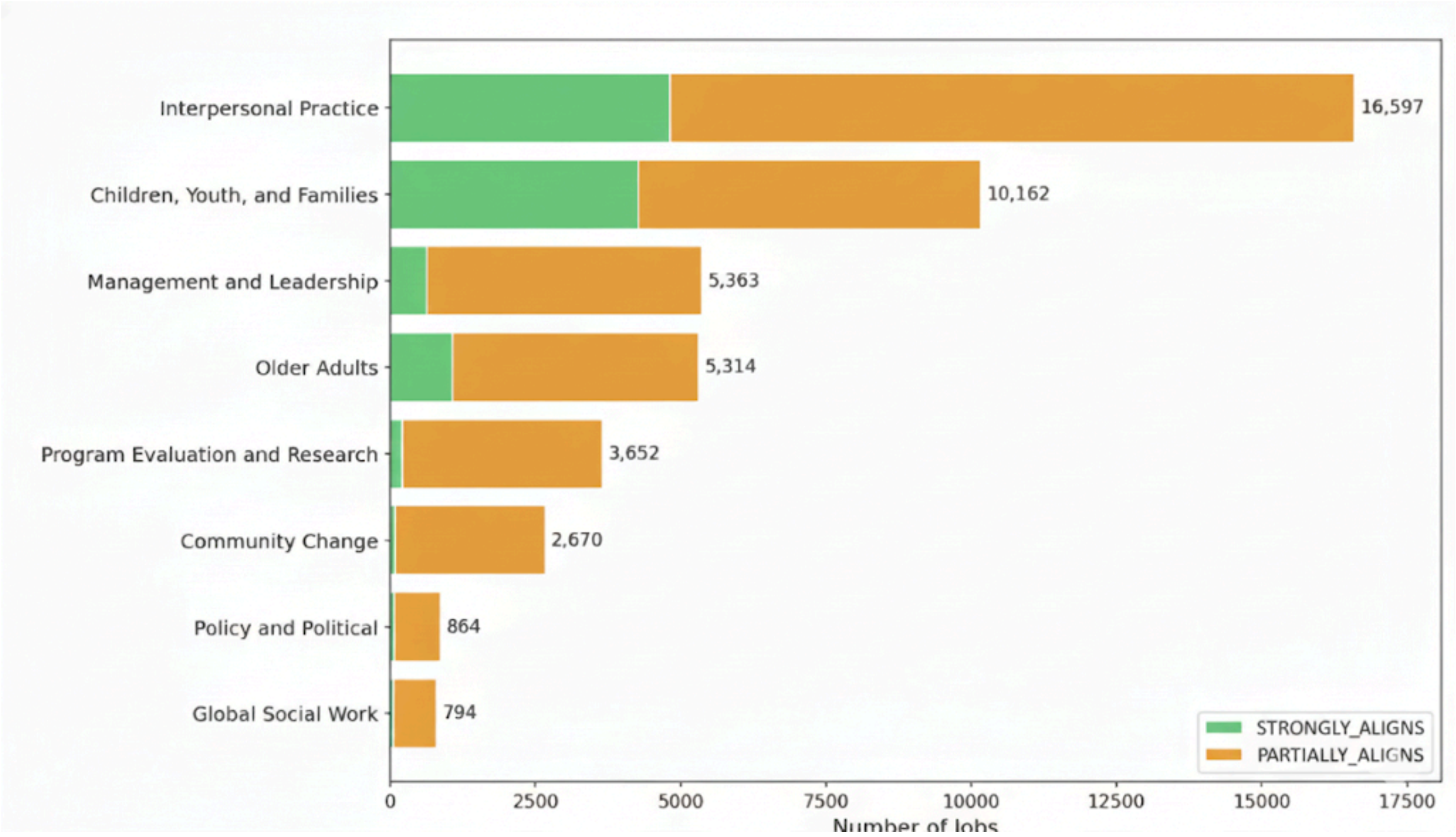

**Figure 2.**

*Coefficients for Specialization Co-occurrence Patterns*

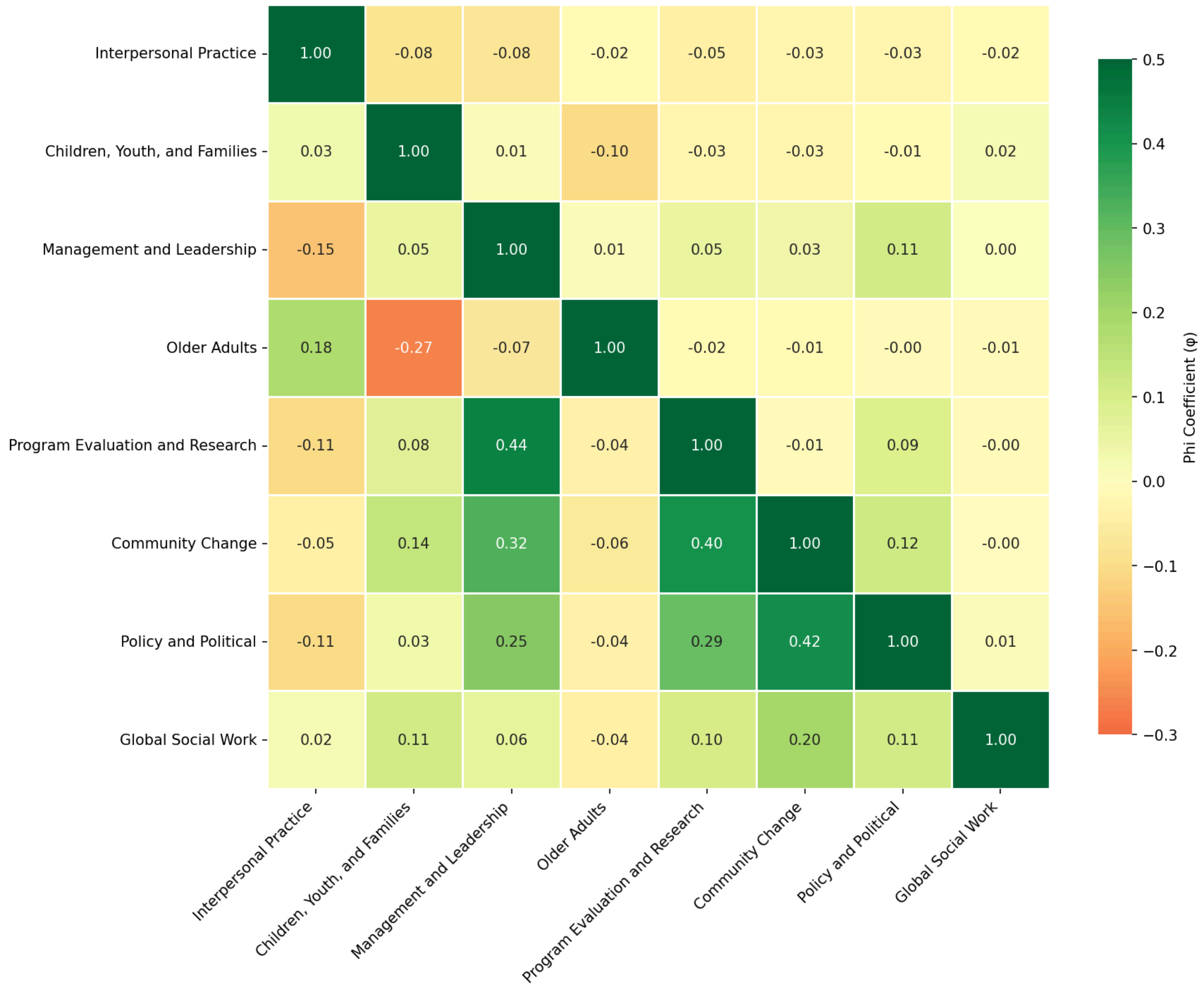

*Note.* Lower triangle displays phi coefficients for all aligned positions (n = 23,732); upper triangle displays phi coefficients for strongly aligned positions only (n = 7,791).Values near zero indicate specializations rarely co-occur; positive values indicate specializations frequently appear together in job postings.

# Appendix A

## Search Terms

| | | |
|---|---|---|
| Addiction Therapist | Addiction Specialist | Advocacy Coordinator |
| Behavioral Health | Behavioral Health Clinician | Care Coordinator |
| Case Manager | Child Protective Services | Child Welfare |
| Child Welfare Worker | Clinical Social Worker | Community Organizer |
| Community Outreach Specialist | Correctional Social Worker | Counselor |
| Crisis Counselor | Data Analyst | Discharge Planner |
| Evaluation Coordinator | Family Preservation Specialist | Family Therapist |
| Forensic Social Worker | Foster Care Case Manager | Geriatric Social Worker |
| Grant Writer | Grants Manager | Healthcare Case Manager |
| Homeless Services Coordinator | Hospice Social Worker | Human Services |
| Human Services Analyst | Licensed Clinical Social Worker | Medical Social Worker |
| Mental Health Professional | Mental Health Therapist | Nonprofit Manager |
| Oncology Social Worker | Outpatient Therapist | Palliative Care Social Worker |
| Patient Navigator | Policy Analyst | Program Analyst |
| Program Coordinator | Program Evaluation | Program Evaluator |
| Program Manager | Program Officer | Psychotherapist |
| Quality Improvement Specialist | Refugee Immigration Services | Research Analyst |
| Research Associate | Research Coordinator | School Social Worker |
| Social Impact | Social Service | Social Welfare |
| Social Work | Social Worker | Substance Abuse Counselor |
| Therapist | Veterans Services | Youth Services Coordinator |

## Appendix B

## Specialization Definitions

| **Specialization** | **Core Focus** |
|---|---|
| Interpersonal Practice | Clinical direct practice with individuals, families, groups; mental health; substance abuse; evidence-based interventions |
| Children, Youth, and Families | Child welfare; foster care; adoption; school social work; juvenile justice; family preservation |
| Management and Leadership | Nonprofit administration; program management; supervision; strategic planning; fiscal oversight |
| Older Adults | Gerontological practice; healthcare; hospice; long-term care; discharge planning |
| Program Evaluation and Research | Research design; data analysis; outcomes measurement; needs assessment; implementation science |
| Community Change | Community organizing; coalition building; advocacy campaigns; grassroots mobilization |
| Policy and Political | Policy analysis; legislative advocacy; government relations; think tanks |
| Global Social Work | Refugee/immigrant services; international development; humanitarian aid; human rights |